\documentclass[aps,preprint,nofootinbib]{revtex4}%
\pdfoutput=1
\usepackage[utf8]{inputenc}
\usepackage{amsfonts}
\usepackage{amsmath}
\usepackage{amssymb}
\usepackage{float}
\usepackage{graphicx}%
\providecommand{\U}[1]{\protect\rule{.1in}{.1in}}

\begin{document}
\title{Quasinormal modes and black hole hairs in AdS}
\author{Mariano Chernicoff$^1$, Gaston Giribet$^2$, Julio Oliva$^3$, Ricardo Stuardo$^3$}

\affiliation{\small{$^1$Departamento de F\'isica, Facultad de Ciencias, Universidad Nacional Aut\'onoma de M\'exico
A.P. 70-542, CDMX 04510, M\'exico}}
\affiliation{\small{$^2$Physics Department, University of Buenos Aires and IFIBA-CONICET,
Ciudad Universitaria, pabell\'on 1, 1428, Buenos Aires, Argentina.
}}
\affiliation{\small{$^3$Departamento de F\'isica, Universidad de Concepci\'on, Casilla, 160-C, Concepci\'on, Chile.}}

\begin{abstract}
Holography relates the quasinormal modes frequencies of AdS black holes to the pole structure of the dual field theory propagator. These modes thus provide the timescale for the approach to thermal equilibrium in the CFT. Here, we study how such pole structure and, in particular, the time to equilibrium can get modified in the presence of a black hole hair. More precisely, we consider in AdS a set of relaxed boundary conditions that allow for a low decaying graviton mode near the boundary, which triggers an additional degree of freedom. We solve the scalar field response on such background analytically and non-perturbatively in the hair parameter, and we obtain how the pole structure gets affected by the presence of a black hole hair, relative to that of the usual AdS black hole geometry. The setup we consider is a massive 3D gravity theory, which admits a one-parameter family deformation of BTZ solution and enables us to solve the problem analytically. The theory also admits an AdS$_3$ soliton which gives a family of vacua that can be constructed from the hairy black hole by means of a double Wick rotation. The spectrum of normal modes on the latter geometry can also be solved analytically; we study its properties in relation to those of the AdS$_3$ vacuum.

\end{abstract}

\maketitle

\newpage
\section{Introduction}

Three-dimensional gravity \cite{3D} has proven to be a fruitful arena to investigate different aspects of the AdS/CFT correspondence \cite{Malda} that, in other setups, could hardly be addressed. Such is the case of the microscopic computation of non-supersymmetric black holes entropy \cite{Strominger} or the study of possible resolutions of the information loss paradox \cite{Eternal}. Gravity in $D=3$ dimensions has also shown to be useful to explore generalization of the holographic correspondence to other scenarios, including de-Sitter space \cite{dSCFT}, warped deformations of AdS space \cite{WAdSCFT}, and asymptotically flat spaces \cite{BMSCFT}. It has also provided manageable toy models to investigate holography in higher-spin theories \cite{HS1, HS2}, in higher-derivative theories \cite{CG}, and in non-relativistic theories \cite{Lifshitz}. Exact results in quantum gravity like the computation of the partition function of pure Einstein theory \cite{Maloney, Yin}, cf. \cite{Acosta}, are also an example of how far the application of AdS/CFT to $3D$ gravity can lead us: There have been speculation about solving the theory holographically even at finite central charge $c$ \cite{Witten, Ising}, the regime in which the gravity theory is highly quantum.

One of the features of AdS/CFT that can be addressed in $3D$ with more details than in higher dimensions is the problem of relaxing the boundary conditions: Different boundary conditions in AdS generically leads to different dual CFTs, with different field content and different unitarity properties \cite{LogGravity}. Nevertheless, {\it some} features of these different CFTs may be shared among them, as they may be associated to particular aspects of the bulk theory that are indistinguishable. This enables one to speculate about the possibility of finding a consistent geometric dual description for a broader class of CFTs, probably including non-unitary ones \cite{Vafa}. The latter possibility is particularly interesting as in $D=2$ dimensions there exist non-unitary CFTs that are of importance in physics \cite{c=01, c=02, c=03, c=04, c=05}.

As a matter of fact, $3D$ gravity has already been considered as dual to non-unitary CFTs: In \cite{Log1, Log2, Log3, Log4}, bulk computations in massive $3D$ gravity \cite{NMG} have been shown to reproduce the correct form of stress-tensor correlation functions of a logarithmic CFT \cite{LogCFT}, which is non-unitary. In this paper, we will study such a model of massive $3D$ gravity, but at a different point of the parameter space. We will consider the so-called New Massive Gravity (NMG) \cite{NMG}, which is a parity-even theory of $3D$ gravity that has very interesting properties. In particular, it contains a rich phase space of black holes, including black holes with a softly decaying hair in AdS$_3$ \cite{OTT, NMG2}. The hair parameter, which will be denoted $b$, controls the asymptotic boundary conditions of AdS$_3$ space, switching between the standard Brown-Henneaux AdS$_3$ boundary condition \cite{BH} when $b=0$ and a weakened AdS$_3$ asymptotic \cite{OTT} when $b\neq 0$. This will enable us to study how some properties of the dual CFT$_2$ change when the asymptotic AdS$_3$ boundary conditions get relaxed. Specifically, we will focus our attention on the quasinormal modes (QNM) computation on the hairy black hole background.

Black holes in AdS holographically correspond to an state of the dual field theory that is approximately thermal, and the decay of the field in the black hole background corresponds, in the boundary, to the decay of perturbations of such thermal state. Quasinormal modes then provide the timescale for the approach to thermal equilibrium. This means that the quantization conditions coming from the QNM computation in the AdS$_3$ black hole geometry are in correspondence with the pole structure of the propagator in the dual CFT$_2$ at finite temperature. More precisely, the retarded Green function in the momentum space can be obtained from the relative coefficient of two independent solutions of a probe field on the black hole background, after the appropriate boundary conditions have been imposed. In the $3D$ case this can be computed explicitly \cite{QNM}, for example for the BTZ black hole \cite{BTZ}, which obeys the standard boundary conditions in AdS$_3$. Here, by explicitly computing the QNM of the hairy black hole, we will investigate how the pole structure of the boundary retarded propagator gets modified in presence of a gravitational hair that distorts the asymptotically boundary conditions in AdS$_3$.

But, before being more specific about the gravity theory that we will consider, let us summarize the main properties of its dual CFT$_2$:
\begin{itemize}
\item Its central charge is given by $c=3\ell/G$, where $\ell $ is the AdS$_3$ radius and $G$ is the $3D$ Planck length. In other words, $c$ is twice the central charge of the CFT$_2$ that is dual to pure Einstein gravity \cite{BH}, with the factor of 2 being due to the massive graviton contribution.
\item Its field content will also differ from that of the CFT$_2$ that is dual to GR, the reason being that the boundary conditions we will consider here present a weaker fall-off near the boundary, implying a different normalization condition for the fields.
\item Like other theories of its sort, NMG suffers from the so-called bulk/boundary unitarity clash, meaning that the sign of $G$ for which the bulk excitation have positive energy is the one for which the central charge of the dual CFT$_2$ is negative. This implies that for a well-defined bulk theory with local degrees of freedom, the dual CFT will be non-unitary, cf. \cite{ZDG, MMG, EMG}.
\item AdS/CFT correspondence still works: Cardy formula exactly reproduces the entropy of the hairy black holes in the bulk \cite{Nosotros}. This is analogous to what happens in Einstein gravity, where the entropy of BTZ black holes can be written as the Cardy formula of a CFT$_2$ with a Brown-Henneaux central charge \cite{Strominger}. However, since the hairy black holes have non-constant curvature, the CFT$_2$ derivation of their entropy is even more remarkable, as it cannot be deduced from the usual arguments of \cite{Soda, Saida}, cf. \cite{StromingerCardy}.
\item The asymptotic isometries with relaxed boundary conditions form two copies of Virasoro algebra, and the associated Noether charges yield the right central charge, $c$. This shows that local conformal symmetry is compatible with the relaxed boundary conditions \cite{OTT}. In addition, the theory exhibits an extra current, which is studied in \cite{ComingSoon}.
\item The quantization conditions that follow from the QNM computation get modified by the presence of the gravitational hair. In other words, the relaxed boundary conditions plus the backreaction of the hair on the black hole geometry change the quantization condition that would holographically correspond to the pole structure of the dual theory propagator.
\end{itemize}

Here, we will be involved with the latter problem: We will compute the QNM of the hairy black hole explicitly and show that it deviates from the structure of the BTZ black hole. As we will see, special features of the QNM spectrum in the BTZ black hole, like the property of having equispaced imaginary part, get affected in the presence of low decaying gravitational modes. We will also solve analytically the NM frequencies on the so-called AdS$_3$ soliton, which may also be connected to the QNM of the hairy black hole as both geometries are related by a double Wick rotation. The paper is organized as follows: In Section 2, we introduce the hairy black hole solution of massive $3D$ gravity we consider. In Section 3, we study the causal structure and the stability conditions of the solutions. In Section 4, we compute the QNM spectrum of a scalar probe in the hairy black hole background and perform a comparative analysis with the usual AdS$_3$ black hole computation. In Section 5, we consider the AdS$_3$ soliton background, for which we also compute the normal modes analytically. Section 6 contains our final remarks.


\section{The setup}

We consider the three-dimensional parity-even massive gravity theory known as New Massive Gravity (NMG) \cite{NMG}. This theory, which exhibits many attractive features, is defined by the action
\begin{equation}
S=\frac{1}{16\pi G}\int d^3x\sqrt{-g}\Big{[}R-2\lambda +\frac{1}{m^2} K \Big{]}\ ,
\end{equation}
where
\begin{equation}
K=R_{\mu\nu}R^{\mu\nu}-\frac{3}{8}R^2.
\end{equation}
Newton's constant gives the $3D$ Planck length $\ell _P = G$; hereafter we set the convention $\ell _P = 1/4$. $\lambda $ is the cosmological constant, and $m$ is a mass parameter associated to the graviton mass: In fact, this theory propagates a massive spin-2 field that, at linearized level, is described by the Fierz-Pauli theory. The field equations derived from this action read
\begin{equation}\label{FEs}
G_{\mu\nu}+\lambda g_{\mu\nu}-\frac{1}{2m^{2}}K_{\mu\nu}=0\ ,
\end{equation}
where $K_{\mu\nu}$ is a symmetric rank-2 tensor, quadratic in the curvature and of fourth order in the metric, whose trace coincides with $K$. Field equations (\ref{FEs}) admit very interesting black hole solutions with different asymptotics. This is one of the reasons why NMG is attractive to investigate AdS/CFT correspondence and its ramifications.

For generic values of the couplings, NMG theory has two maximally symmetric
solutions, with two different effective cosmological constants. These two vacua, however, coincide when $\lambda=m^{2}$, and the curvature radius of the
AdS$_3$ solution in this case reads $\ell^{2}=-\frac{1}{2m^{2}}$. Also at this particular point, the family of stationary asymptotically AdS$_3$ black hole solutions get enhanced and admits a one-parameter generalization of the BTZ geometry. This can be thought of as the black holes in AdS$_3$ admitting a low decaying gravitational hair. The theory also admits other static asymptotically AdS$_3$; among them, an AdS$_3$ soliton \cite{OTT}; see below.

The metric of the static hairy black hole is given by%
\begin{equation}
ds^{2}=-\left(  r^{2}+br-\mu\right)  dt^{2}+\frac{dr^{2}}{\left(r^{2}+br-\mu \right)} +r^{2}d\phi^{2}\ , \label{bh}%
\end{equation}
where we have set $\ell =1$. The range for the coordinates here is taken to be $-\infty<t<\infty$, $0\leq \phi \leq\pi$, $0\leq r<\infty$. The parameter $b\in \mathbb{R}$ can be regarded as a gravitational hair, and $\mu$ is an integration constant related to the ADT mass \cite{DeserADT1, DeserADT2}; see (\ref{massa}) below. The functional relation between these two parameters and the event and Cauchy horizons, $r_+$ and $r_-$ respectively, is given by
\begin{equation}
b=-(r_++r_-), \qquad \mu=-{r_+r_-}.
\end{equation}
Equivalently,
\begin{align}\label{rmasrmenos}
r_{+}  &  =\frac{1}{2}\left(  -b+\sqrt{b^{2}+4\mu}\right)  \ ,\\
r_{-}  &  =\frac{1}{2}\left(  -b-\sqrt{b^{2}+4\mu}\right)  \ .
\end{align}
Notice that the inner Cauchy horizon exists provided $b<0$ and $-\frac{b^{2}}{4}<\mu<0$. Alternatively, we can think of $b$ as a control parameter that allows us to depart from the BTZ black hole geometry. This interpretation will be useful in order to interpret the results that we will present below. It is important to emphasize that the black hole described by (\ref{bh}) is asymptotically AdS$_3$ in a way that is weaker than the standard asymptotic of, say, the BTZ solution \cite{OTT}; for example, the next-to-leading behavior of the $g_{tt}$ component of the metric is of order $\mathcal{O}(r)$. Besides, unlike BTZ, spacetime (\ref{bh}) with $b\neq 0$ has non-constant scalar curvature; namely
\begin{equation}
R=-6-\frac{2b}{r}\ \ .
\end{equation}
Actually, the curvature diverges at $r=0$. Figure 1 summarizes the black hole causal structures as a function of the parameter $b$ and the corresponding ADT mass $M$, where
\begin{equation}
M=\mu +\frac{b^2}{4}\ \ .\label{massa}
\end{equation}
Here we are using $G=1/4$. Throughout this letter we will use either $\mu$ or $M$, depending on which gives a clearer physical picture, and even for simplicity in some expressions we will use both with the understanding that $\mu=\mu(M)$.
\begin{figure}[h]
\begin{center}
\includegraphics[width=15cm]{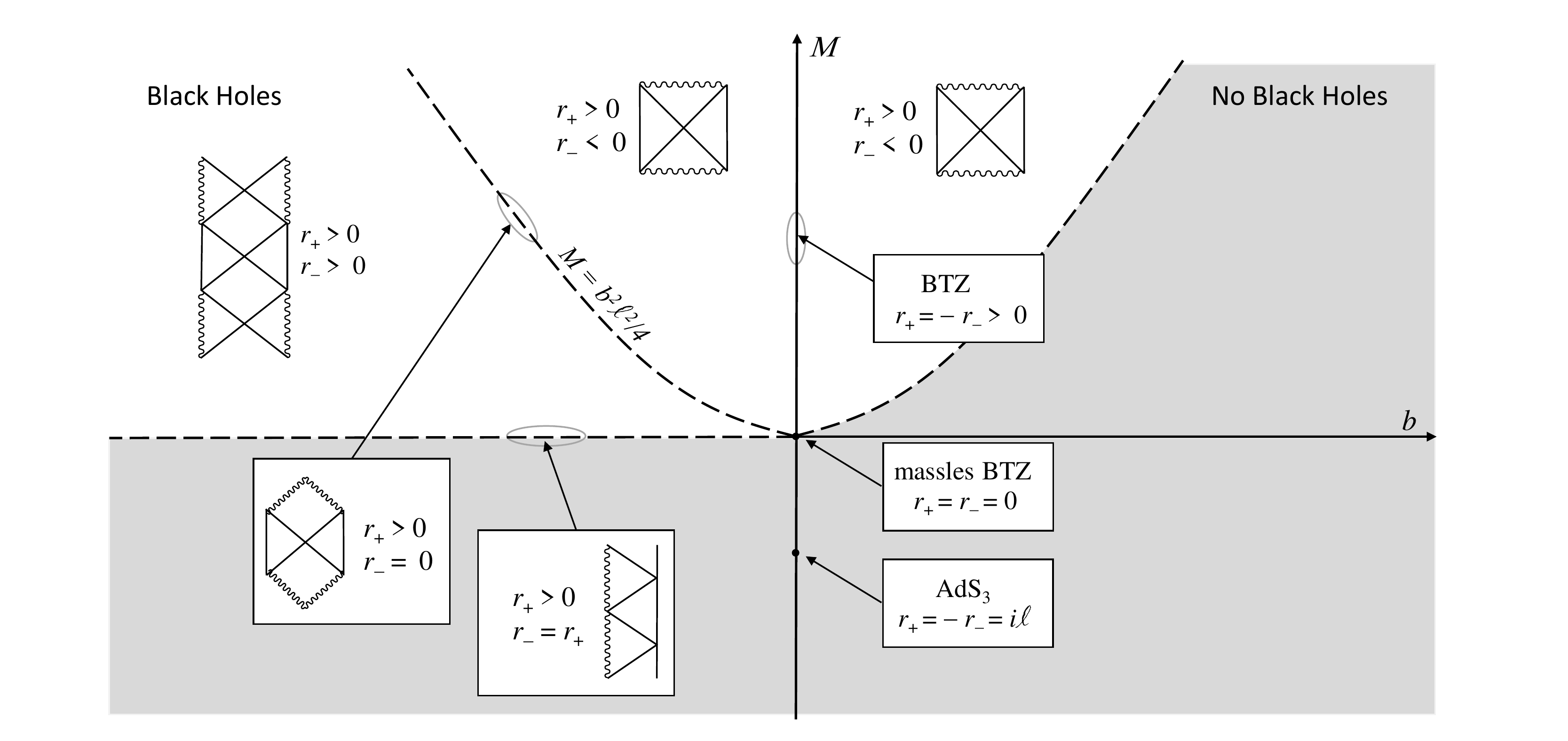}
\caption{Spectrum of hairy black holes in NMG}
\end{center}
\end{figure}

Now, let us say a few words about other solution we mentioned before, the one to which we referred as the AdS$_3$ soliton: Its metric can be written as
\begin{equation}
ds^{2}=-\left(  a+\cosh\rho\right)  ^{2}dt^{2}+d\rho^{2}+\sinh^{2}\rho
d\phi^{2}\ , \label{soliton}%
\end{equation}
where $-\infty<t<\infty$, $0\leq \phi \leq\pi$, $0\leq \rho<\infty$, and where $a>-1$. The latter choice for the range of $a$ ensures that the solution is regular: The Ricci scalar associated to geometry (\ref{soliton}) reads
\begin{equation}
R_{sol}=-\frac{2a+6\cosh\rho}{a+\cosh\rho}\ .
\end{equation}
The parameter $a$ can be thought of as a measure of how much this geometry deviates from global AdS$_3$, the latter space corresponding to $a=0$. This gravitational soliton is related to the hairy black hole (\ref{bh}) through a double Wick rotation \cite{OTT}, and in Section 5 we will make use of this local equivalence to relate the QNM calculation in both geometries, after the corresponding boundary conditions are properly identified.

\section{Causal structures and stability}

Let us study the scalar field response on the hairy black hole background (\ref{bh}). Consider a massive, non-minimally coupled scalar probe on that geometry, then, the corresponding equation of motion is given by
\begin{equation}
\left(  \square-m^{2}-\xi R\right)  \Phi=0\ .
\end{equation}
We use ingoing Eddington-Finkelstein coordinates, $v=t+r_{\ast}$ with
$dr_{\ast}=f\left(  r\right)  ^{-1}dr$ and  $f\left(  r\right)  =r^{2}+br-\mu
$, and we consider the separable ansatz%
\[
\Phi(v,r,\phi)=r^{-\frac{1}{2}}\Psi(r)e^{-i\omega v}e^{in\phi}\ ,
\]
where $\omega $ represents the frequency associated to the null direction $v$, and $n$ is the quantized angular momentum. This yields
\begin{equation}
\frac{d}{dr}\left(  f(r)\frac{d\Psi}{dr}(r)\right)  -2i\omega\frac{d\Psi}%
{dr}(r)-V(r)\Psi(r)=0\ ,\label{Psi}%
\end{equation}
with the effective potential having the form
\begin{equation}
V(r)=m^{2}+\frac{\left(  1-8\xi\right)  \left(  b+3r\right)  }{4r}+\frac
{1}{4r^{2}}\left(  4n^{2}+\mu\right)  \ .
\end{equation}
This potential simplifies when $\xi=1/8$, which in the massless case leads to
the conformally coupled scalar probe, and enables for an
explicit integration of the QNM. Hereafter, for simplicity, we will focus on this value of the non-minimal coupling $\xi $. In this case one
is left with a potential that does not depend on the hair parameter $b$ explicitly; however, there is an implicit dependence as the possible values of $\mu$ compatible with the existence of an event
horizon is actually $b$-dependent. As shown in \cite{HoroHube},
having demanded regularity of the solution at the horizon, and Dirichlet boundary condition at infinity imply that $\operatorname{Im}\left(  \omega\right)  <0$, provided the effective
potential is positive definite in the domain of outer communications. In fact,
\begin{equation}
\int_{r_{+}}^{\infty}dr\left(  f\left(  r\right)  |\Psi^{\prime}|^{2}+V\left(
r\right)  |\Psi|^{2}\right)  =-\frac{|\omega|^{2}|\Psi\left(  r_{+}\right)
|^{2}}{\operatorname{Im}\omega}\ .
\end{equation}
Notice that, as usual, the angular momentum of the scalar field, $n$, contributes to the stability of the perturbation. For $m^{2}\geq0$, the stability of the
s-wave is guaranteed if $\mu\geq0$.

The spaces described by metric (\ref{bh}) can be characterized in terms of their ADT mass $M$ and the value of the intensity of the hair $b$. In such case, and for a suitable normalization of Newton's
constant, the lapse function reads (see also \cite{ComingSoon})%
\begin{equation}
f\left(  r\right)  =\left(  r+\frac{b}{2}\right)  ^{2}-M\ .
\end{equation}
Figure 1 depicts all possible causal structures in terms of $M$ and $b$. The proof of stability that we have given above applies to all the black holes that lie from the half-parabola $4M=b^{2}$ with $b<0$ to the right. These black holes are characterized by possessing a single horizon. For black holes
in the complementary region, the previous stability analysis is not conclusive.
The equation for the QNM in all these backgrounds can be
solved in closed form, and below we study the spectra for different values
of $M$ and $b$.

\section{Quasinormal frequencies}

Both the hairy black hole and the AdS$_3$ soliton can be shown to be conformally flat. Therefore, it is natural to study the dynamics of a conformally coupled scalar field as a probe. The corresponding field equation is
\begin{equation}
\left(  \square-\frac{1}{8}R\right)  \Phi=0\ .
\end{equation}

Let us be reminded of the fact that the Ricci scalar of these metrics is not constant, and therefore the conformal coupling does not simply represent an effective mass, but a rather complicated function that can even change its sign in the bulk. $R$ does approach a finite value at infinity, though: near infinity, $-\frac{1}{8}R\rightarrow m^{2}=-\frac{3}{4}$, and this means that the {\it asymptotic} effective mass is above the Breitenlohner-Freedman (BF) bound. This leads to the two possible asymptotic behaviors
$r^{-\Delta_{\pm}}$ with $\Delta_{+}=3/2$ and $\Delta_{-}=1/2$. Both of these branches
are actually normalizable; however, since here we are interested in Dirichlet boundary conditions, the coefficient in front of the branch that goes like $r^{-1/2}$ is set to zero. Notice that these boundary conditions are AdS$_3$ invariant.

As usual, we consider a separation ansatz%
\begin{equation}
\Phi\sim e^{-i\omega t+in\phi}H\left(  r\right)  \ ,\label{ansatztr}%
\end{equation}
and we begin by considering the scalar probe on the black hole geometry. It is convenient to define a new coordinate $x$ such that%
\begin{equation}
r=\frac{r_{+}r_{-}\left(  1-x\right)  }{r_{-}-r_{+}x}\ ,
\end{equation}
which maps $r_{+}<r<+\infty$
to $0<x<\frac{r_{-}}{r_{+}}$. Notice that this change of coordinates is
valid for both possible signs of $r_{-}$, i.e. both in presence and in absence
of the inner Cauchy horizon, provided $r_{-}\neq0$. It turns out that the case $r_{-}=0$ has to be analyzed separately; and the same happens with the extremal case $r_{+}=r_{-}$.

Let us first study the solution near the horizon: The equation for $H\left(  z\right)  $ can be easily integrated in terms of
the hypergeometric functions $_{2}F_{1}$ (hereafter denoted $F$), and, after imposing ingoing boundary condition at the horizon, one is left with%
\begin{equation}
H\left(  z\right)  \sim\left(  2b(1-x^{2})\sqrt{M}+(1+x^{2})b^{2}%
+2\mu(1+x)^{2}\right)  ^{\frac{1}{4}}x^{\frac{-i\omega}{2\sqrt{M}}}(1-x)^{-\frac{in}{\sqrt{\mu}}}F\left(  a_{1},b_{1},c_{1},x\right)  \ ,
\end{equation}
where the symbol $\sim$ means up to a constant of integration and
\begin{equation}
a_{1}=\frac{-2in+\sqrt{\mu}}{2\sqrt{\mu}}\ ,\  b_{1}=\frac{\left( \mu-2in\right)  \sqrt{M}-2i\omega\sqrt{\mu}}{2\sqrt{\mu M
}}\ ,\  c_{1}=\frac{\sqrt{M}-i\omega}{\sqrt
{M}}\ .\label{params}%
\end{equation}

The expansion near infinity, on the other hand, reads%
\begin{equation}
H\left(  z\right)  \sim\left(  \frac{r_{-}}{r_{+}}-x\right)  ^{\frac{1}{2}%
}\left[  A+B\left(  \frac{r_{-}}{r_{+}}-x\right)  +\mathcal{O}\left(  \left(
\frac{r_{-}}{r_{+}}-x\right)  ^{2}\right)  \right]
\end{equation}
where the symbol $\sim$ has the same interpretation as above and
\begin{equation}
A=F\left(  a_{1},b_{1},c_{1},\frac{r_{-}}{r_{+}}\right) \ \ \text{ and
}\ \ B=F^{\prime}\left(  a_{1},b_{1},c_{1},\frac{r_{-}}{r_{+}}\right)  \ .
\end{equation}
Notice that in the asymptotic region, $(  \frac{r_{-}}{r_{+}}-x)
\sim r^{-1}$; therefore, $H\left(  x\left(  r\right)  \right)  $ actually behaves
as expected, namely $\sim r^{-\Delta_{\pm}}$ with $\Delta_{+}=3/2$ and $\Delta_{-}=1/2$. As mentioned above, this is because the spacetime is asymptotically AdS$_3$ and the conformal coupling of the scalar produces an effective mass term at infinity yielding $m_{\text{conf}}^{2}=-3/4$, which is above the BF bound. Imposing Dirichlet boundary conditions, this results in%
\begin{equation}
F\left(  a_{1},b_{1},c_{1},\frac{b+2\sqrt{M}}{b-2\sqrt{M}%
}\right)  =0\ .\label{polecondition}%
\end{equation}
This equation defines the spectrum of QNM on the hairy black
hole. Therefore, the retarded Green function in the dual theory in momentum
space reads
\begin{equation}
G_{\text{ret}}\left(  \omega,n\right)  =-\frac{B}{A}=-\frac{F^{\prime}\left(
a_{1},b_{1},c_{1},\frac{r_{-}}{r_{+}}\right)  }{F\left(  a_{1},b_{1}%
,c_{1},\frac{r_{-}}{r_{+}}\right)  }\ ,
\end{equation}
with $a_{1}$, $b_{1}$ and $c_{1}$ defined in (\ref{params}). That is,
\begin{equation}
G_{\text{ret}}\left(  \omega,n\right)  =-\frac{F^{\prime}\left(
\frac{1}{2}-\frac{in}{\sqrt{M-b^2/4}}
,
\frac{1}{2}-\frac{in}{\sqrt{M-b^2/4}}-\frac{i\omega }{2\sqrt{M}}
,
1-\frac{i\omega }{\sqrt{M}}
,\frac{r_{-}}{r_{+}}\right)  }{F\left(
\frac{1}{2}-\frac{in}{\sqrt{M-b^2/4}}
,
\frac{1}{2}-\frac{in}{\sqrt{M-b^2/4}}-\frac{i\omega }{2\sqrt{M}}
,
1-\frac{i\omega }{\sqrt{M}}
,\frac{r_{-}}{r_{+}}\right)  }\ ,
\end{equation}
The quantized quasinormal
frequencies thus correspond to the poles conditions in the retarded Green function. In the BTZ limit, i.e. when the hair parameter $b$ tends to zero, one gets%
\begin{equation}
F\left(  \frac{-2in+r_{+}}{2r_{+}},\frac{r_{+}-2in-2i\omega}{2r_{+}}%
,\frac{2r_{+}-2i\omega}{2r_{+}},-1\right)  \sim\frac{\Gamma\left(
2-\frac{i\omega}{r_{+}}-\frac{2in}{r_{+}}\right)  }{\Gamma\left(  \frac{3}%
{4}+\frac{in}{2r_{+}}-\frac{i\omega}{2r_{+}}\right)  \Gamma\left(  \frac{3}%
{4}-\frac{in}{2r_{+}}-\frac{i\omega}{2r_{+}}\right)  }=0\ ,
\end{equation}
which, as expected, reduces to the quasinormal frequencies of a conformally coupled scalar field on BTZ geometry \cite{QNMBTZ}; namely%
\begin{equation}
\omega_{\text{BTZ}}=\pm \, n-i\, r_{+}\left(  \frac{3}{2}+2p\right)  \ ,\label{btzfreqs}%
\end{equation}
with the mode number being $p\, =\, 0,\, 1,\, 2, \, ..$. For the BTZ black hole, the imaginary part of the
frequency, being linear in the mode number, leads to an equispaced damping spectrum. The s-wave modes are all located on the imaginary axis, and the imaginary part
of the frequencies is always negative, confirming the stability of the
propagation. Also, for BTZ the real part of the frequency is independent of the mass $M_{\text{BTZ}}=r_{+}^{2}$. As we show below, this is not the case for the hairy black hole with $b\neq 0$, for which both the oscillation frequencies and the damping depend on the mass. This is one of the features that get affected by the presence of the low decaying gravitational hair.

In order to find the quasinormal frequencies of the hairy black hole, we need to solve the field equation (\ref{polecondition}). Since for $b\neq 0 $ this is a trascendental equation, we have to solve it numerically: The frequencies always come in pairs with the same damping but opposite sings for the real part. Therefore, the plots below have to be understood as having a specular image of the frequencies in the quadrant $\operatorname{Im}\omega<0$,
$\operatorname{Re}\omega<0$. Qualitatively different
behaviors are exhibited by the different types of black holes that appear in Figure 1.

Before exploring the spectra of such solutions explicitly, it is worth rewriting the equation for
the radial dependence of the scalar field in a Schr\"{o}dinger-like form
\begin{equation}
-\frac{d^{2}\tilde{H}}{dr_{\ast}^{2}}+V\left(  r\left(  r_{\ast
}\right)  \right)  \tilde{H}=\omega^{2}\tilde{H}\ ,
\end{equation}
where $\tilde{H}(r)=\sqrt{r}H(r)$ and
the tortoise coordinate being defined by $dr_{\ast}=(r-r_{+})^{-1}(r-r_{-})^{-1}\, dr$, maps $r_{+}<r<+\infty$ to
$-\infty<r_{\ast}<0$. The potential can be written explicitly in terms of $r_{\ast}$, yielding
\begin{equation}
    V(r_{\ast})=\frac{(4n^2+\mu)M}{\left( 2\sqrt{M}\cosh\left(r_{\ast}\sqrt{M} \right) + b\sinh\left(r_{\ast}\sqrt{M} \right)\right)^{2}}\ , \label{EffectivePotential}
\end{equation}
which, in the BTZ case, i.e. $b=0$, (\ref{EffectivePotential}) reduces to a P\"{o}schl-Teller potential which vanishes on the horizon and approaches a constant near the boundary  (see e.g. \cite{Cooper}). Interestingly, in the presence of the gravitational hair, $b\neq 0$ introduces a deformation, which can be mapped back to the P\"{o}schl-Teller form by a complex shift in the radial coordinate. It is also important to notice that when $4M=b^2$ ($\mu=0$) and $n=0$, that is, for the s-wave when the black hole singularity is null, the problem reduces to that of a free particle on the semi-infinite line. As shown below, these facts have an imprint on the spectrum of the conformal scalar probe.

\subsection{Black holes with a single horizon}

In Section 2 we provided a general argument about the stability of the scalar
probes on black holes that applies to those with a single horizon $r_{+}$
($r_{-}\leq0$), which correspond to black holes represented by points inside the parabola $4M=b^{2}$ in Fig. 1. Fig. 2 shows different spectra that depart
continuously from their corresponding BTZ counterparts (in bullets). It is worth noticing that, due to the hair parameter $b$, both the real and imaginary parts of the frequencies do depend on the black hole mass, however, when $b$ vanishes $\operatorname{Re} \omega $ does not (see \eqref{btzfreqs}). Notice also that for this family of black holes, $b$ is bounded. In particular, when $-4M^2<b$, by black hole configurations that hide a null singularity, and when $b<4M^2$, by geometries with naked singularities. There is a clear non-analyticity of the spectra as a function of $b$ at the BTZ point $b=0$. For negative values of the hair parameter the s-wave modes are purely damped.
\begin{figure}[h!]\label{n0nullsing}
\begin{center}
\minipage{0.45\textwidth}
  \includegraphics[width=\linewidth]{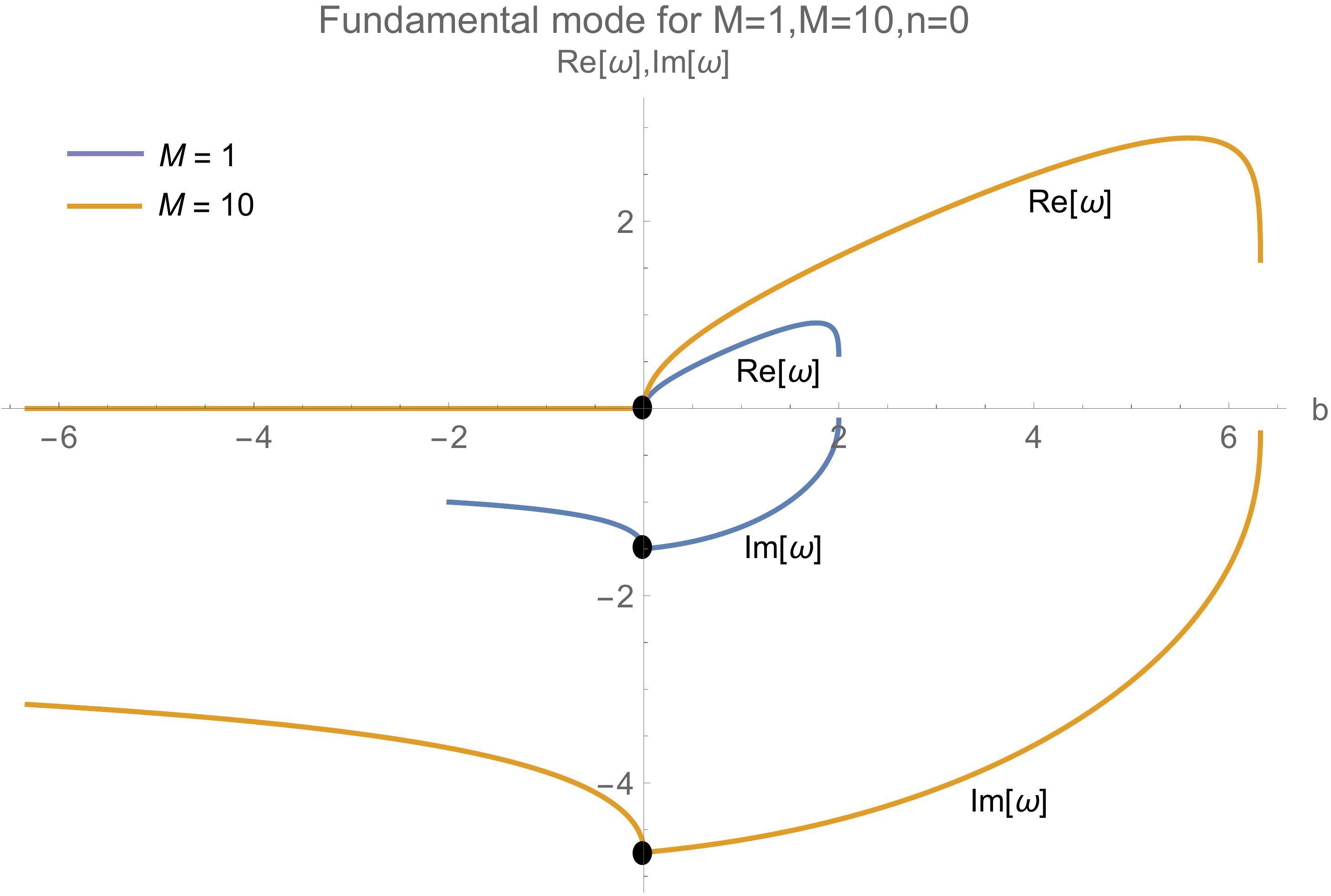}
\endminipage \ \
\minipage{0.45\textwidth}
  \includegraphics[width=\linewidth]{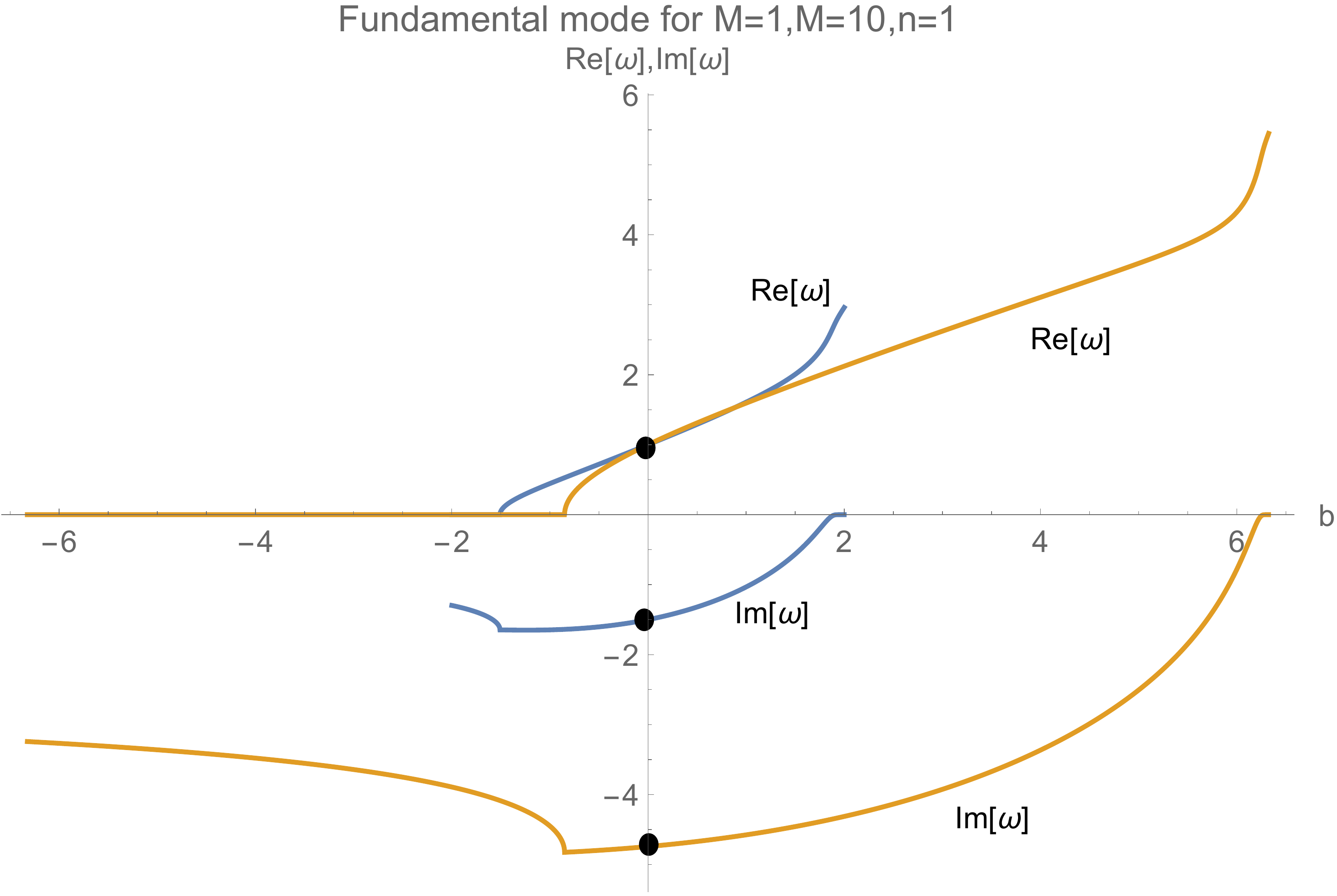}
\endminipage
\caption{Spectra for fundamental modes of the scalar field on the black holes with $-b<4M^2<b$. The spectra on these black holes depart continuously from the BTZ spectrum of fundamental modes $\omega_{BTZ}=n-\frac{3}{2}\sqrt{M}i$, in bullets.}
\end{center}
\end{figure}

\subsection{Black hole surrounding a null singularity}

As it can be seen from Fig. 1, for negative values of the hair parameter $b$, according to (\ref{rmasrmenos}), if
$4M=b^{2}$ the event horizon is located at $r=|b|=r_{+}$ and surrounds a null
singularity located at $r=0$. Now, it is interesting to notice
that the s-wave of the conformal scalar probe, namely (\ref{ansatztr}) with
$n=0$, can be integrated in terms of radicals, leading to%
\begin{equation}\label{nullsingsolution}
R\left(  r\right)  = C_{1}\left(  r-r_{+}\right)  ^{-\frac{i\omega}{r_{+}}%
}r^{\frac{i\omega}{r_{+}}-\frac{1}{2}}+C_{2}\left(  r-r_{+}\right)
^{\frac{i\omega}{r_{+}}}r^{-\frac{i\omega}{r_{+}}-\frac{1}{2}}\ .
\end{equation}
Imposing ingoing conditions at the event horizon requires setting $C_{2}=0$. On the other hand, expanding the remaining part as $r$ becomes large, one can actually see that  (\ref{nullsingsolution}) allowed $r^{-1/2}$ and $r^{-3/2}$
behaviors. With this in mind, one should notice that it is impossible to
impose Dirichlet boundary conditions on this solution because in doing so, it would require that $C_{1}=0$, and therefore
$R\left(  r\right)  =0$. Then, one is forced to conclude that the black holes with lapse function $f(r)=r(r-r_+)$ do not support spherically symmetric QNM for conformally coupled scalar fields. In other words, spherically symmetric modes cannot be excited on these black holes if Dirichlet boundary condition are imposed\footnote{It is interesting to notice that for a backreacting conformally coupled scalar field, NMG does admit a black hole with the same causal structure \cite{Mokhtar,Eloy} and the scalar field at infinity behaves as $r^{-1/2}$.}. The situation changes drastically when one introduces angular momentum. The equation for the conformal scalar probe in that case can be integrated and, after imposing ingoing boundary conditions, one obtains
\begin{equation}
R\left(  r\right)  \sim r^{-\frac{1}{2}}\, I_{\nu}\left(  -\frac{2n}{r_{+}}%
\sqrt{1-\frac{r_{+}}{r}}\right)  \ ,
\end{equation}
where $\nu=-2i\omega/r_{+}$ and where $I_{\nu}$ stands for the modified Bessel
function of the first kind. Notice that in the asymptotic region, the argument
of the Bessel function goes to a finite number, where the function does not
have any singular behavior. Therefore, the asymptotic expansion of the field in
this case reads%
\begin{equation}
R\left(  r\right)  \sim r^{-\frac{1}{2}}\left[  I_{\nu}\left(  -\frac
{2n}{r_{+}}\right)  +I_{\nu}^{\prime}\left(  -\frac{2n}{r_{+}}\right)
\frac{1}{r}+\mathcal{O}\left(  r^{-2}\right)  \right]  \ .
\end{equation}
Dirichlet boundary conditions thus imply that the frequencies must be
quantized according to%
\begin{equation}
I_{-2i\omega/r_{+}}\left(  -\frac{2n}{r_{+}}\right)  =0\ .
\end{equation}
Figure 3 depicts the spectra for the fundamental and first three excited modes ($n=1$) with the real and imaginary parts of the frequencies as a function of the horizon radius. We see a non-analytic behavior for a critical value of $r_+$ on each curve, above which, the real part of the frequencies vanishes and the scalar field perturbations are purely damped. Modes with higher angular momentum, $n>1$, have a similar behavior.
\bigskip
\begin{figure}[h!]\label{n0nullsing}
\begin{center}
\minipage{0.45\textwidth}
  \includegraphics[width=\linewidth]{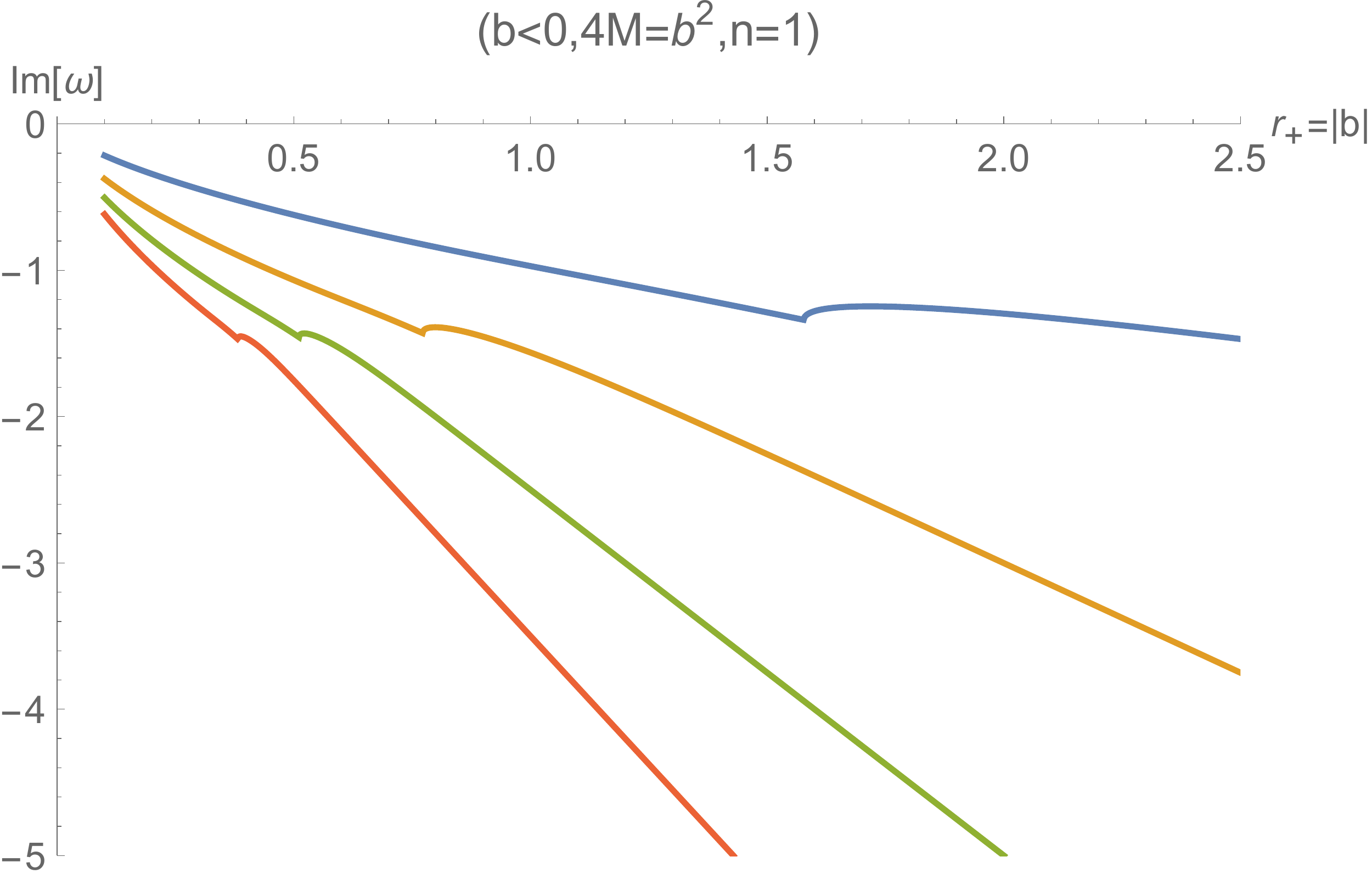}
\endminipage \ \
\minipage{0.45\textwidth}
  \includegraphics[width=\linewidth]{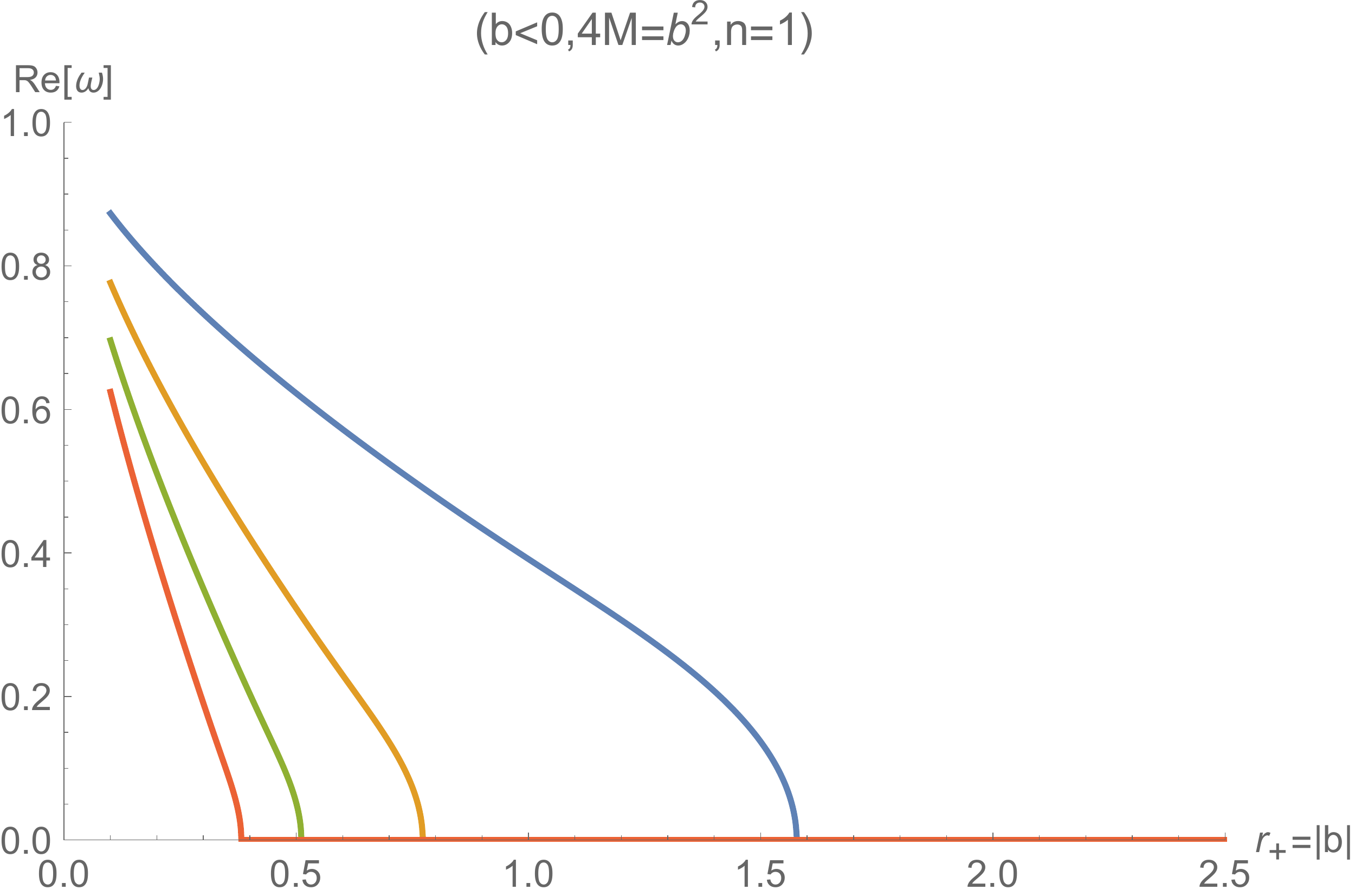}
\endminipage
\caption{Spectra for the scalar field on the black hole with a null singularity, with $n=1$. Blue color depicts the fundamental modes, and yellow, green and blue correspond to the first, second and third harmonics, respectively.}
\end{center}
\end{figure}

\subsection{Black holes with event and Cauchy horizons}

When both $r_{-}$ and $r_{+}$ are positive, the causal structure of these hairy black holes changes and a Cauchy horizon appears inside the event horizon. These solutions are not perturbativelly connected to the BTZ branch. Still, the quantization condition for the
frequencies comes from equation (\ref{polecondition}). Figure 4 shows an example of  such spectra. The frequencies are purely damped, regardless of the value of the angular momentum of the scalar field.
\begin{figure}[h!]\label{extremalspectra}
\begin{center}
 \includegraphics[scale=0.4]{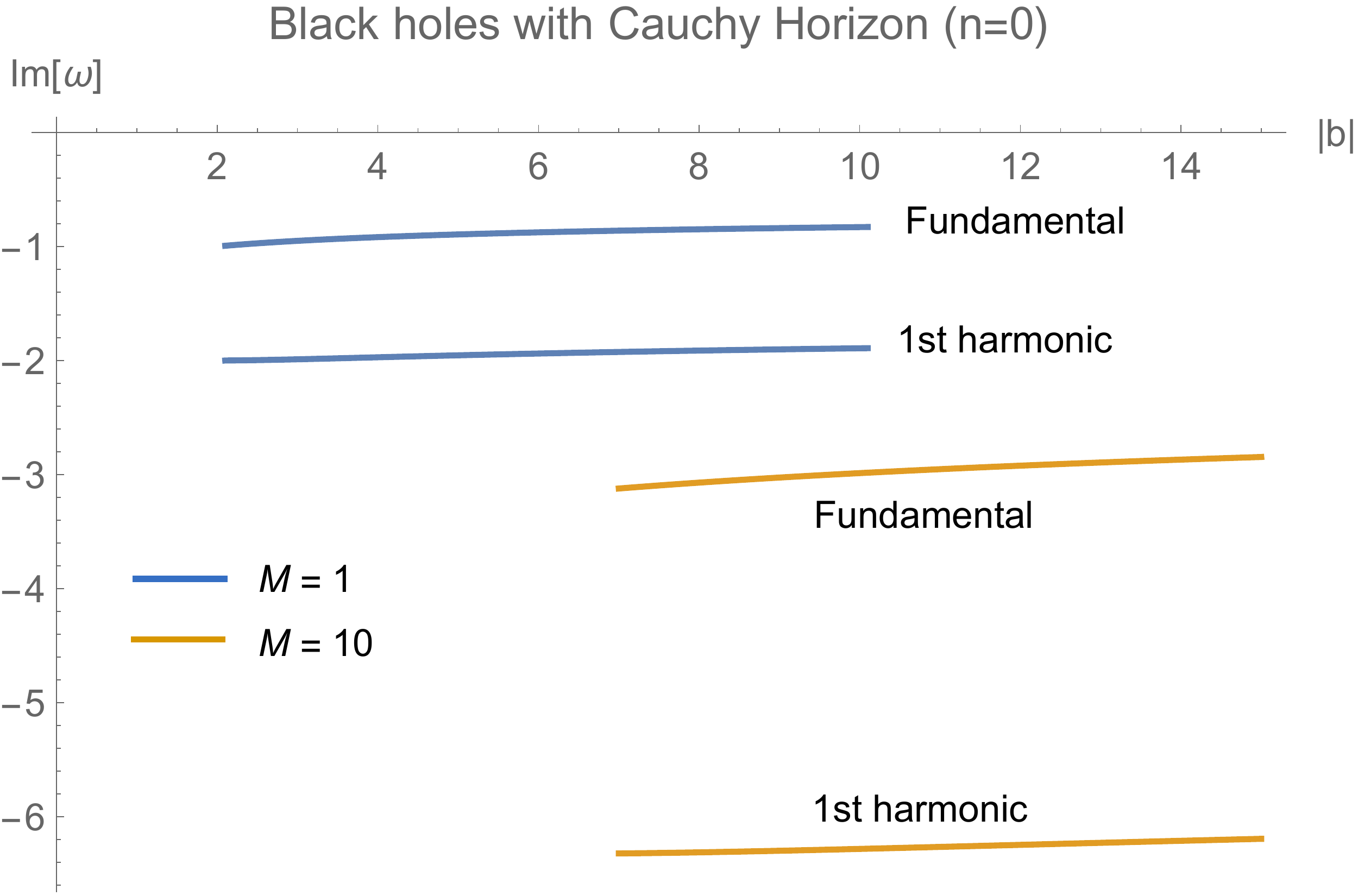}
\caption{Example spectra from black holes with Cauchy horizons as a function the hair parameter. The real part vanishes even when there is angular momentum on the scalar field.}
\end{center}
\end{figure}

\subsection{The extemal case}

When the hair parameter $b$ is negative, the black hole with $\mu =0$ turns out to be extremal, with $r_{+}=r_{-}=|b|/2$. In this case, the conformal scalar equation can be solved in terms of confluent hypergeometric functions, and the branch that is analytic at
the horizon, in Schwarzschild-like coordinates, reads $R\left(  r\right)  \sim
e^{\frac{i\omega}{r-r_{+}}}$. The analyticity of this branch can be checked, as usual, in terms of the coordinate $v=t+r_{*}$, which leads to a regular metric on the future horizon. Connecting this branch with Dirichlet
boundary conditions at infinity, namely $R\left(  r\right)  \sim r^{-3/2}$,
leads to the following quantization condition:%
\begin{equation}\label{espectroextremo}
K_{-\frac{n}{r_{+}}}\left(  -\frac{i\omega}{r_{+}}\right)  =0\ ,
\end{equation}
where $K_{\nu}\left(  x\right)  $ is a modified Bessel function of the second kind. It is worth mentioning that, as for the black holes surrounding a null singularity, the case without angular momentum does not allow to impose Dirichlet boundary condition at infinity, while the inclusion of such allows to find non-vanishing modes fulfilling \eqref{espectroextremo}, with a very interesting structure (see Fig. 5).

For a given $n>0$, one has the fundamental as well as a finite number of overtones. As the radius of the horizon increases, the number of allowed overtones is diminished until a critical radius of the black hole horizon above which, again it is impossible to fulfil the boundary conditions.
\begin{figure}[h!]\label{extremalspectra}
\begin{center}
 \includegraphics[scale=0.4]{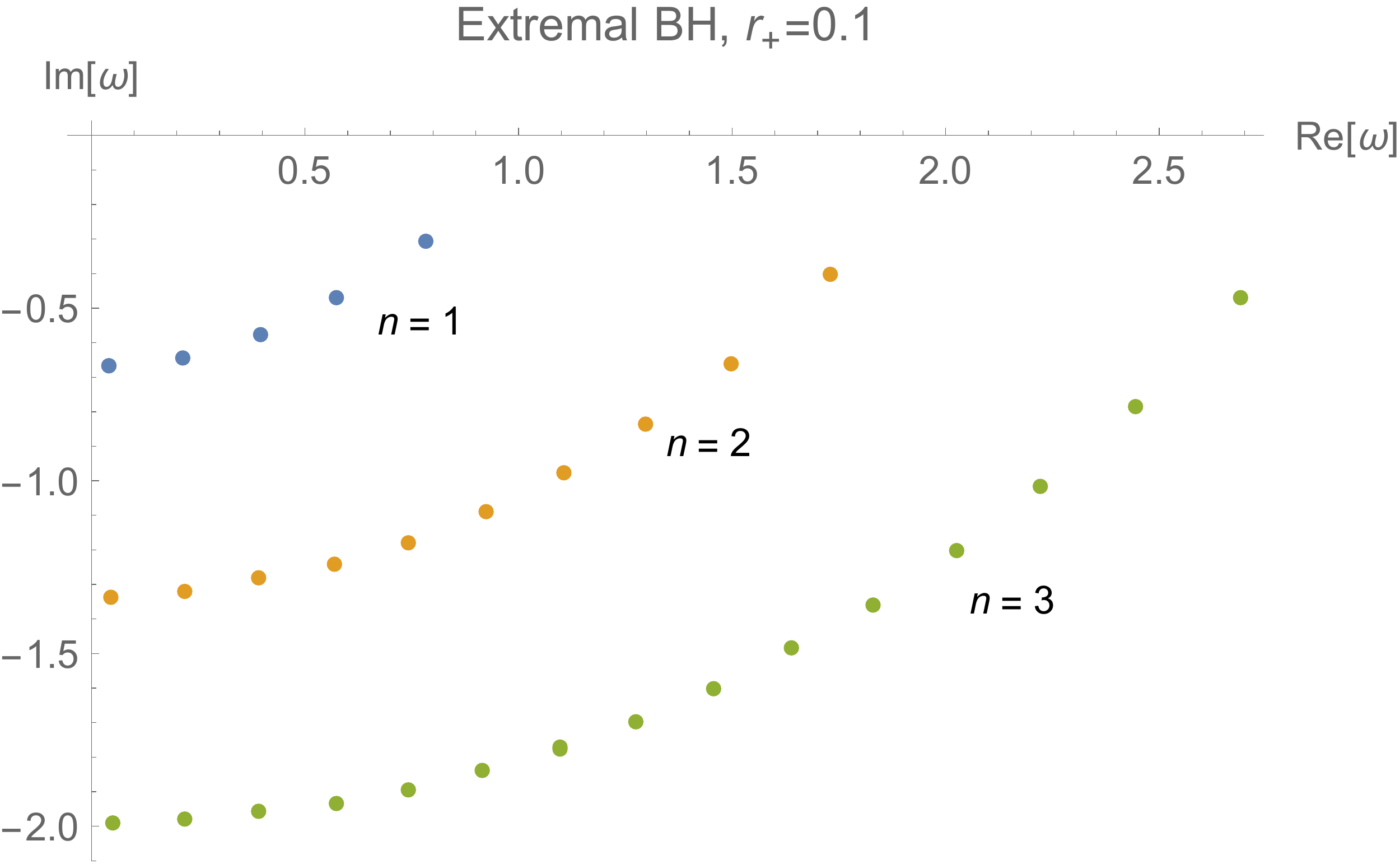}
\caption{Spectra for the extremal black holes with $r_+=0.1$, for different values of the angular momentum of the scalar probe. As $r_+$ increases the higher overtones near the vertical axis stop being valid modes and eventually, above a critical value of $r_+$ all the modes for a given $n$ are forbidden. Such critical value for $n=1$ is near $r_+\sim 0.66$. It is possible to check that the effective mass induced by the non-minimal coupling in the near horizon $AdS_2$ geometry, it is equal to the two-dimensional BF bound.}
\end{center}
\end{figure}

Following the same approach as we did to obtain equation (\ref{EffectivePotential}), the potential for the Schr\"{o}dinger equation in this case takes the simpler form of a shifted centrifugal potential
\begin{equation}
    V(r_{\ast})=\frac{1}{4}\frac{(4n^2-r_{+}^{2})}{\left(r_{+}r_{\ast}-1\right)^{2}},
\end{equation}
with $-\infty <r_{\ast}<0$. One can check that this expression can be obtained by taking the $M\rightarrow 0$ limit in \eqref{EffectivePotential}.

\section{Normal modes on the soliton}

As mentioned earlier, besides the black holes presented above, NMG also admits static asymptotically AdS$_3$ solutions which represents AdS$_3$ solitons. In what follows, we will study the response of the conformal scalar probe on the geometry given by (\ref{soliton}).

Let us consider the separable ansatz%
\begin{equation}
\Phi\sim e^{-i\omega t+in\phi}R\left(  \rho\right)  \ ,
\end{equation}
and define the change of coordinate given by %
\begin{equation}
x=\left(  \frac{a-1}{a+1}\right)  \frac{\cosh\rho-1}{\cosh\rho+1}\ ,
\end{equation}
which maps $0<\rho<+\infty$ to $0<x<\frac{a-1}{a+1}$. This change of coordinates is not valid if $a=1$, so we will study this case separately. The branch that
respects regularity at the origin leads to (up to an integration constant)%
\begin{equation}
R\left(  x\right)  \sim x^{n/2}\sqrt{(x-1)a+x+1}(1-x)^{\frac{\omega}%
{\sqrt{1-a^{2}}}}F\left(  a_{2},b_{2},c_{2},x\right)  \ ,
\end{equation}
with%
\begin{equation}
a_{2}=\frac{2\omega+\sqrt{1-a^{2}}}{2\sqrt{1-a^{2}}}\ ,\ b_{2}=\frac
{\sqrt{1-a^{2}}+2\omega+2n\sqrt{1-a^{2}}}{2\sqrt{1-a^{2}}}\ ,\ c_{2}=1+n\ .
\end{equation}
The behavior at the boundary (i.s. $\rho$ going to infinity) up to a constant of integration is given by
\begin{equation}
R\left(  x\right)  \sim\left(  \frac{a-1}{a+1}-x\right)  ^{1/2}\left[
\tilde{A}+\tilde{B}\left(  \frac{a-1}{a+1}-x\right)  \right]  \ ,
\end{equation}
where
\begin{equation}
\tilde{A}=F\left(  a_{2},b_{2},c_{2},\frac{a-1}{a+1}\right) \ \  \text{ and
}\ \ \tilde{B}=F^{\prime}\left(  a_{2},b_{2},c_{2},\frac{a-1}{a+1}\right)  .
\end{equation}
Dirichlet boundary condition implies
\begin{equation}
F\left(
\frac{1}{2} +\frac{\omega }{\sqrt{1-a^2}}
,
\frac{1}{2} +\frac{\omega }{\sqrt{1-a^2}}+n
,
1+n
,\frac{a-1}{a+1}\right)  =0\ . \label{qsol}%
\end{equation}
This equation provides the normal frequencies of the conformal scalar on the
soliton as a function of the parameter $a$ and being an algebraic transcendental equation it has to be solved numerically. Notice that in the AdS$_3$ limit, one
has $a=0$, and therefore%
\begin{equation}
F\left( \frac{1}{2} + \omega,\frac{1}{2}+\omega+n,1+n,-1\right)  \sim
\frac{\Gamma\left(  2+2\omega+n\right)  }{\Gamma\left(  \frac{3}{4}%
+\frac{\omega}{2}+\frac{n}{2}\right)  \Gamma\left(  \frac{3}{4}-\frac{\omega
}{2}+\frac{n}{2}\right)  }=0\ ,
\end{equation}
which leads to the correct, fully resonant spectrum for a conformal scalar
probe on AdS.

In order to study the special case $a=1$, it is convenient to define the following change of coordinate
\begin{equation}
    z=\frac{\cosh(\rho)-1}{\cosh(\rho)+1}\  ,
\end{equation}
which maps $0<\rho<\infty $ to $0<z<1$. The branch that respects regularity at the origin reads
\begin{equation}
    R(z) \sim \sqrt{1-z}\, J_{n}(\sqrt{z}\, \omega),
\end{equation}
where $J_{n}$ is the Bessel function of the first kind. The corresponding asymptotic expansion at infinity (i.e. $\rho\rightarrow\infty$) takes the form
\begin{equation}
    R(z) \sim (1-z)^{1/2}\left[ \bar{A}+\bar{B}(1-z) \right] ,
\end{equation}
where
\begin{equation}
    \bar{A} = J_{n}(\omega) \ \  \text{ and }\ \ \bar{B}= J'_{n}(\omega).
\end{equation}
Then, imposing Dirichlet boundary condition leads to
\begin{equation}
    J_{n}(\omega)=0;
\end{equation}
which means that, when $a=1$, the normal frequencies are simply given by $\omega = \alpha_{n,p}$ where $\alpha_{n,p}$ is the $p$-th zero of the $n$-th Bessel function. The quantization condition therefore leads to the results shown in Fig. 6.
\begin{figure}[!h]
\begin{center}
\minipage{0.45\textwidth}
  \includegraphics[width=\linewidth]{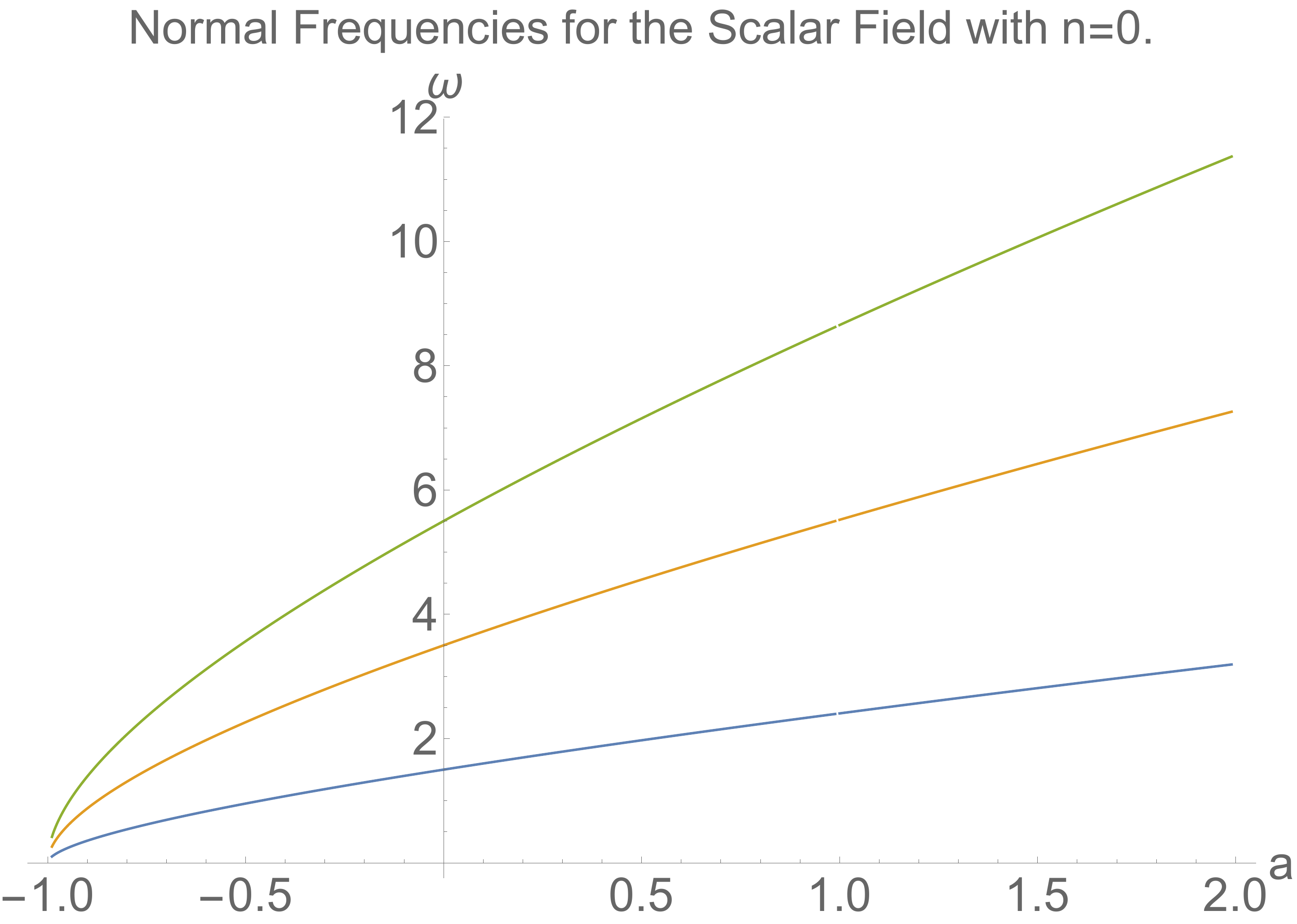}
\endminipage \ \
\minipage{0.45\textwidth}
  \includegraphics[width=\linewidth]{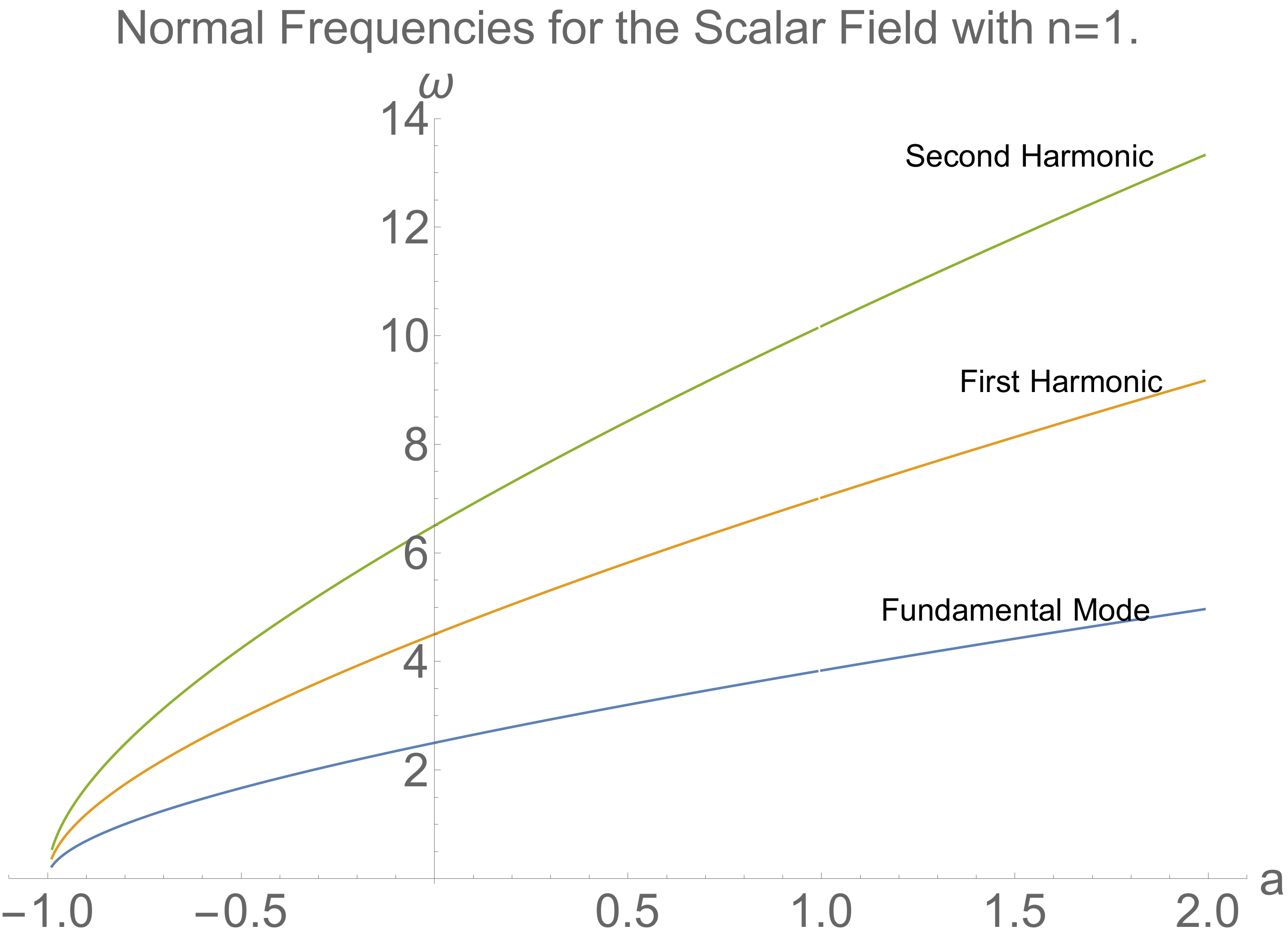}
\endminipage
\caption{Spectra for the scalar field on the soliton with two different values of $n$. The left plot corresponds to a scalar field with no angular momentum while the plot on the right $n=1$.}
\end{center}
\end{figure}

It is worth noticing that when rewriting the radial equation for the soliton in a Schr\"{o}dinger-like form, the potential takes a remarkably simple expression. Namely, the equation reads%
\begin{equation}
-\frac{d^{2}\tilde{R}}{d\rho_{\ast}^{2}}+V\left(  \rho\left(  \rho_{\ast
}\right)  \right)  \tilde{R}=\omega^{2}\tilde{R}\ ,
\end{equation}
where $\tilde{R}=\sinh^{1/2}\rho$ $R$ and $d\rho_{\ast}=d\rho/\left(
a+\cosh\left(  \rho\right)  \right)  $, with the potential%
\begin{align}
V\left(  \rho_{\ast}\right)   &  =\frac{1}{4}\frac{\left(  4n^{2}-1\right)  \left(
a^{2}-1\right) }{\sinh^{2}(\rho_{\ast}\sqrt{a^2-1})}\text{ for }a>1\ \text{, }0<\rho_{\ast
}<\frac{2\tanh^{-1}\left[  \sqrt{\frac{a-1}{a+1}}\right]  }{\sqrt{a^{2}-1}}\\
V\left(  \rho_{\ast}\right)   &  =\frac{1}{4}\frac{4n^{2}-1 }{\rho_{\ast}^2}\text{ for }a=1\ \text{, }0<\rho_{\ast
}<1\\
V\left(  \rho_{\ast}\right)   &  =\frac{1}{4}\frac{\left(  4n^{2}-1\right)  \left(
1-a^{2}\right)  }{\sin^{2}(\rho_{\ast}\sqrt{1-a^2})}\text{ for }-1<a<1\ \text{, }0<\rho_{\ast}<\frac{2\tan
^{-1}\left[  \sqrt{\frac{1-a}{1+a}}\right]  }{\sqrt{1-a^{2}}}\ .
\end{align}


Before we conclude our study for the NM of the scalar field on (\ref{soliton}), it is worth saying a few words regarding the close relation between the QNM frequencies given by equation (\ref{polecondition}) and those obtained from (\ref{qsol}).  As mentioned before, the AdS$_3$ soliton can be obtained by performing a double Wick rotation on (\ref{bh}).
More precisely, given
\begin{equation}
ds^{2}=-\left(  r^{2}+br-\mu\right)  dt_{BH}^{2}+\frac{dr^{2}}{r^{2}+br-\mu
}+r^{2}d\phi_{BH}^{2}\ ,
\end{equation}
and defining
\begin{equation}
t_{BH}=i\tilde{\phi}\, ,\ \ \phi_{BH}=i\tilde{t}\, ,\ \  r=-\frac{b}{2}%
+\sqrt{\frac{b^{2}}{4}+\mu}\cosh\rho \ ,
\end{equation}
leads to%
\begin{equation}
ds^{2}=-\left(  \frac{b^{2}}{4}+\mu\right)  \left(  -\frac{b}{\sqrt{b^{2}%
+4\mu}}+\cosh\rho\right)  ^{2}d\tilde{t}^{2}+d\rho^{2}+\left(  \frac{b^{2}}%
{4}+\mu\right)  \sinh^{2}\rho d\tilde{\phi}^{2}\ .
\end{equation}
Notice that, since the coordinate $\tilde{t}$ is non-compact, the prefactor in the
$g_{\tilde{t}\tilde{t}}$ component can be absorbed. Also important, to avoid a conical
singularity at $\rho=0$, one requires that the period of $\tilde{\phi}$
must be defined in such way that the corresponding prefactor can be again absorbed to define a new angular
coordinate $\phi_{sol}$ which goes from $0$ to $2\pi$. Therefore, identifying%
\begin{equation}
a=-\frac{b}{2\sqrt{M}}\ ,
\end{equation}
one obtains (\ref{soliton}). Motivated by this result, it is sensible to apply a double Wick rotation at the level of the spectrum in momentum space, obtaining%
\begin{align}
\omega_{\text{BTZ}}  &  =i\sqrt{M} \, n_{sol}\ ,\\
n_{\text{BTZ}}  &  =i\sqrt{M} \, \omega_{sol}\ .
\end{align}
Remarkably, these relations allow to map the equation that defines the
spectrum of the conformal scalar on the hairy black hole \eqref{polecondition} to the condition
that defines the spectrum of the the conformal scalar probe on the soliton \eqref{qsol}.
The same has been observed in black holes with other asymptotics, such as Lifshitz \cite{NuestroLifshitz}, where the spectrum for a massive scalar of
both the black hole as well as the soliton obtained by a Wick rotation can be
obtained in a closed form. The ingoing condition on the horizon naturally maps to the regularity at the origin since $\rho=0$ in the soliton geometry corresponds to $r=r_{+}=-\frac{b}{2}+\sqrt{M}$ in the black hole geometry. The same should happen between the planar AdS black hole and the AdS soliton in arbitrary dimensions \cite{HM}, as well as with the rotating black branes obtained by boosting the planar AdS black holes.

\section{Further comments}

In this paper we studied how the quasinormal modes in the AdS$_3$ black hole background get modified when a low decaying gravitational hair is turned on. This hair satisfies a weakened asymptotic conditions in AdS$_3$, and thus it triggers extra degrees of freedom at the boundary. We managed to solve the scalar field response on such background analytically and non-perturbatively in the hair parameter, and we showed how the quasinormal mode spectrum gets modified in the presence of the hair, relative to the BTZ geometry. In particular, the presence of the hair modifies the timescale for the approach to thermal equilibrium in the dual conformal field theory, leading to a spectrum that, unlike the standard spectrum of BTZ, does not have equispaced dampings and introduce a dependence of the black hole mass in both the real and imaginary parts of the corresponding frequencies. For the hairy black holes that are connected to BTZ spacetime, the presence of the gravitational hair extends the life of the s-wave conformal scalar probes (see Figure 1, left panel).

 There is a smooth geometry related by a double Wick rotation with the black holes, that is parameterized by a single integration constant that measures the departure from global AdS$_3$. We showed that the problem of a conformal scalar probe on the soliton is also solvable in an exact manner, leading to a spectrum of normal modes. For small values of the parameter that characterizes the background, the spectrum differs slightly from that of AdS$_3$, and it is therefore close to a fully resonant spectrum. It would be interesting to study how this departure from the resonant behavior affects the energy transfer between modes when a self interaction on the scalar probe or the backreaction on the geometry are turned on. This energy transfer is enhanced in AdS due to the equispaced property of the spectrum and it leads to the initial departure from AdS in the turbulent cascade that ends in black hole formation in higher dimensions \cite{Bizon:2011gg,Dias:2011ss} .

We proved the stability of the propagation of a probe scalar on the hairy black holes for an arbitrary mass of the scalar $m^2$ when the black hole possesses a single event horizon, and later computed the exact spectrum in the massless, conformally coupled case, which requires a non-minimal coupling with the scalar curvature\footnote{For the rotating, asymptotically flat hairy black holes in NMG, the quasinormal modes for a massless scalar probe where obtained in an exact manner in \cite{Figueroa}.}. The integrability in this case is intrinsically related to the fact that both the hairy black holes and the soliton are conformally flat geometries, which is behind of the exact solvability of a massless fermion \cite{Gonzalez:2014voa}. The conformal factor that maps the hairy asymptotically AdS black holes and solitons to flat space has a singular behavior in the asymptotic region. This implies that even though the spacetime is conformally flat, and the solutions of a conformally coupled source on the hairy black holes could be obtained from their integration on flat space, a non-trivial spectrum emerges due to the presence of the new timelike AdS boundary introduced by the singular conformal factor.

It would also be interesting to see whether more general (Robin) boundary conditions allow to obtain s-wave modes for the black hole surrounding a null singularity as well as in the extremal case. As we showed, such s-wave modes are absent for Dirichlet boundary conditions, but since the conformal coupling with the curvature effectively induces a mass term at infinity that is within the unitarity window, one may explore more general boundary conditions.

\section*{Acknowledgments}
The authors thank Francisco Correa for comments. The work of M.C. is partially supported by Mexico's National Council of Science and Technology (CONACyT) grant A1-S-22886 and DGAPA-UNAM grant IN107520.  The work of G.G. was partially supported by CONICET through the grant PIP 1109-2017. The work of J.O. is supported by FONDECYT grant 1181047. The work of R.S. is funded by CONICYT Fellowships 22191591.  We also thank the Institute of Nuclear Science at UNAM and in particular, Alberto G\"uijosa, for their hospitality during G.G and J.O. visit where this work was started.

  \end{document}